\documentclass[aps,prl,twocolumn,superscriptaddress]{revtex4-2}

\usepackage{graphicx}
\usepackage{xcolor}
\usepackage{units}

\begin{document}

\title{Wetting of two-component drops: Marangoni contraction versus autophobing}

\author{Michiel A. Hack}
\email{m.a.hack@utwente.nl}
\affiliation{Physics of Fluids Group, Faculty of Science and Technology, University of Twente, P.O. Box 217, 7500 AE Enschede, The Netherlands}

\author{Wojciech Kwieci\'{n}ski}
\email{w.kwiecinski@utwente.nl}
\affiliation{Physics of Interfaces and Nanomaterials Group, MESA+ Institute for Nanotechnology, University of Twente, P.O. Box 217, 7500 AE Enschede, The Netherlands}

\author{Olinka Ram\'{i}rez-Soto}
\email{olinka.ramirez@ds.mpg.de}
\affiliation{Max Planck Institute for Dynamics and Self-Organization, Am Fa\ss berg 17, 37077 G\"{o}ttingen, Germany}

\author{Tim Segers}
\affiliation{Physics of Fluids Group, Faculty of Science and Technology, University of Twente, P.O. Box 217, 7500 AE Enschede, The Netherlands}

\author{Stefan Karpitschka}
\affiliation{Max Planck Institute for Dynamics and Self-Organization, Am Fa\ss berg 17, 37077 G\"{o}ttingen, Germany}

\author{E. Stefan Kooij}
\affiliation{Physics of Interfaces and Nanomaterials Group, MESA+ Institute for Nanotechnology, University of Twente, P.O. Box 217, 7500 AE Enschede, The Netherlands}

\author{Jacco H. Snoeijer}
\affiliation{Physics of Fluids Group, Faculty of Science and Technology, University of Twente, P.O. Box 217, 7500 AE Enschede, The Netherlands}

\date{\today}

\begin{abstract}
The wetting properties of multi-component liquids are crucial to numerous industrial applications. The mechanisms that determine the contact angles for such liquids remain poorly understood, with many intricacies arising due to complex physical phenomena, for example due to the presence of surfactants. Here, we consider two-component drops that consist of mixtures of vicinal alkane diols and water. These diols behave surfactant-like in water. However, the contact angles of such mixtures on solid substrates are surprisingly large. We experimentally reveal that the contact angle is determined by two separate mechanisms of completely different nature, namely Marangoni contraction (hydrodynamic) and autophobing (molecular). It turns out that the length of the alkyl tail of the alkane diol determines which mechanism is dominant, highlighting the intricate coupling between molecular physics and the macroscopic wetting of complex fluids. 
\end{abstract}

\maketitle

\begin{figure}[t]
	\begin{center}
		\includegraphics{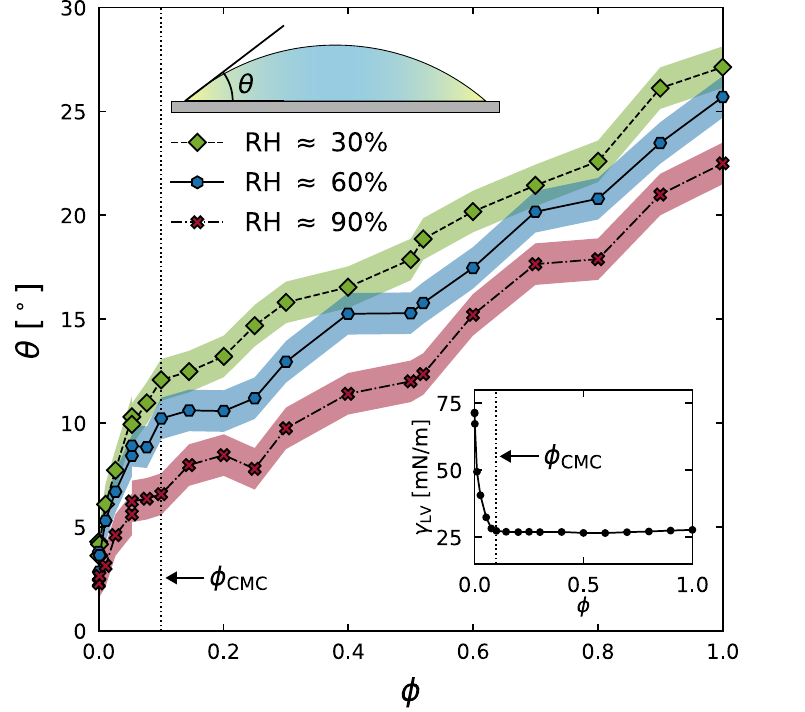}
		\caption{Contact angle ($\theta$) of water--1,2-hexanediol (\mbox{1,2-HD}) mixtures as a function of the mass fraction ($\phi$) of \mbox{1,2-HD}, for various relative humidities (RH). The vertical dotted line indicates the critical micelle concentration ($\phi_\mathrm{CMC}~\approx~0.1$). The filled areas indicate the measurement error.  Schematic: Definition of $\theta$. The mass fraction of \mbox{1,2-HD} (yellow) is higher near the contact line due to selective evaporation. Inset: Surface tension ($\gamma_\mathrm{LV}$) of water--\mbox{1,2-HD} mixtures, measured using the pendant drop method.}
		\label{fig1}
	\end{center}
\end{figure}

Many industrial processes require a fundamental understanding of the wetting properties of liquids on solid surfaces \cite{Starov2018}. Examples are inkjet printing \cite{Wijshoff2010}, oil recovery \cite{Haagh2019}, and lithography \cite{Winkels2011}. A key concept in the description of wetting is the contact angle $\theta$, as defined in Fig.~\ref{fig1}. Properties of the liquid together with the surface chemistry of the solid determine the value of $\theta$ \cite{Young1805, Bonn2009}. The wetting properties and contact angles of single-component liquids have been extensively studied \cite{deGennes2004, Drelich2020}. However, a large number of industrial applications require mixtures of liquids \cite{Leenaars1990} or the addition of surfactant to enhance their spreading properties \cite{Matar2016}. For such complex drops consisting of two or more components, the wetting properties are far from understood. The components may phase separate \cite{Tan2016, Li2018}, selectively evaporate \cite{Sefiane2008}, emulsify \cite{Keiser2017} and adsorb at interfaces \cite{Kim2016}, leading to intricate wetting properties on solid surfaces.  

In this Letter, we study the contact angle $\theta$ of multi-component drops, where one of the phases acts as a surfactant. Figure~\ref{fig1} shows $\theta$ of drops consisting of water--1,2-hexanediol (\mbox{1,2-HD}) mixtures on a piranha solution-cleaned hydrophilic glass substrate (microscope coverslips, Menzel-Gl\"aser) with minimal pinning. The reported angle is attained within seconds after deposition of the drop (cf. Supplementary Material \footnote{See Supplemental Material at WEBSITE LINK\ for additional information on experimental details and a figure showing the attempted collapse the contact angle data using the Marangoni contraction scaling law.}). The key result of Fig.~\ref{fig1} is that $\theta$ continually increases with the \mbox{1,2-HD} mass fraction $\phi$. This is surprising for two reasons. First, \mbox{1,2-HD} has been shown to exhibit surfactant-like properties when mixed with water due to its amphiphilic molecular structure \cite{Haiji1989, Frindi1991, Szekely2007, Tan2018}. Increasing the mass fraction $\phi$ of \mbox{1,2-HD} lowers the surface tension $\gamma_\mathrm{LV}$ (see the inset of Fig.~\ref{fig1}), which normally would lead to an enhanced spreading. However, the opposite trend --reduced spreading-- is found since $\theta$ increases with $\phi$. A second surprise is that this increase continues above the critical micelle concentration $\phi_{\rm CMC}\approx 0.1$  \cite{Romero2007}. Even though $\gamma_\mathrm{LV}$ remains constant for $\phi>\phi_{\rm CMC}$, $\theta$ continues to increase. Here we show that these unexpected features are the result of two mechanisms of different origins -- one of hydrodynamic nature (Marangoni contraction), and the other of molecular nature (autophobing). Additionally, we show to what extent these mechanism are sensitive to the molecular structure of the non-water phase.

\emph{Marangoni contraction.}---We first turn to the hydrodynamic mechanism, which is known as `Marangoni contraction' \cite{Karpitschka2017}. Some multi-component drops (for example water--1,2-propanediol mixtures) can form non-zero contact angles on high-energy surfaces, even though the individual liquids themselves perfectly wet the surface at equilibrium (i.e. $\theta = 0$) \cite{Leenaars1990, Marra1991, Cira2015, Karpitschka2017, Benusiglio2018}. There are two requirements that need to be satisfied for Marangoni contraction to occur: i) one of the two liquids must be significantly more volatile than the other, and ii) the least volatile liquid should have the lowest surface tension of the two liquids. Selective evaporation at the contact line (where the evaporative flux is highest \cite{Deegan1997}) of the volatile component (typically water), then results in a composition gradient in the drop and a surface tension gradient across the drop's interface. This in turn drives a Marangoni flow towards the center of the drop, which opposes the spreading of the drop, such that the drop is `contracted'. The presence of Marangoni contraction invalidates Young's law, which only holds at equilibrium, in the absence of flow \cite{Young1805, deGennes2004}. 

\begin{figure}[t]
	\begin{center}
		\includegraphics{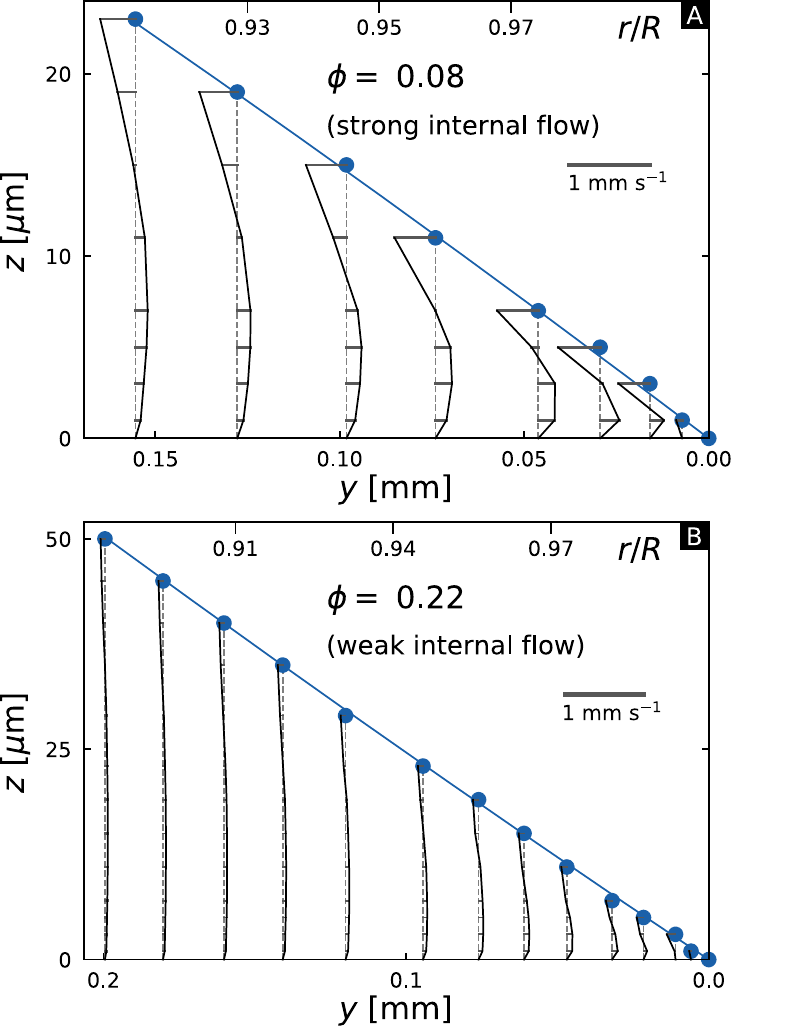}
		\caption{Horizontal velocity component in the drops measured using high resolution micro-particle image velocimetry. The blue line indicates the outer surface of the drop. The horizontal lines indicate the velocity, where the direction is indicated by the location with respect to the vertical dashed line. (a) Velocity field for $\phi = 0.08$ and RH = 71\% ($\theta = 9^\circ$). (b) Velocity field for $\phi = 0.22$ and RH = 40\% ($\theta = 14^\circ$), which is significantly weaker than that in (a).} 
		\label{fig2}
	\end{center}
\end{figure}

Water--\mbox{1,2-HD} mixtures are expected to contract, since \mbox{1,2-HD} is considerably less volatile than water \cite{Li2018}, and has a surface tension lower than that of water (see the inset of Fig.~\ref{fig1}). Figure~\ref{fig2}(a) shows the flow field inside a $\phi = 0.08$ drop, as measured using high resolution micro-particle image velocimetry (experimental details in Supplementary Material \cite{Note1}). The blue line indicates the outer surface of the drop, and the contact line is located at $y=0$. A strong inward flow exists near the surface of the drop, while an outward flow towards the contact line is observed in the bulk of the drop. This flow field is typical for Marangoni contracted drops \cite{Karpitschka2017}. To further test the hypothesis that the increase of $\theta$ is due to Marangoni contraction, we experimentally varied the relative humidity (RH). A low RH enhances the evaporation that drives the flow inside the drop \cite{Semenov2014}. Indeed, Fig.~\ref{fig1} shows that with a lower RH the raise of $\theta$ is indeed significantly enhanced. Furthermore, the scaling with RH as proposed by Karpitschka et al. \cite{Karpitschka2017} shows good agreement with our data in the range $\phi < 0.1$, as shown in the Supplementary Material. Therefore, we conclude that Marangoni contraction is responsible for the enhanced contact angle for water-\mbox{1,2-HD} mixtures at small $\phi$. 

Marangoni contraction alone, however, cannot explain the full range of data in Fig.~\ref{fig1}. At $\phi=1$ all surface tension gradients are removed, but nevertheless, a large (non-zero) $\theta$ is observed. Additionally, a monotonic increase of $\theta$ with $\phi$ is observed in Fig.~\ref{fig1}, even though a decrease in $\theta$ is expected for $\phi \gtrsim 0.6$ due to smaller surface tension gradients and weaker internal flow \cite{Karpitschka2017, Benusiglio2018}. Figure~\ref{fig2}(b) shows the velocity field in a drop at $\phi~=~0.22$, which is almost one order of magnitude smaller than the velocity in the $\phi~=~0.08$ drop (note the difference in length of the horizontal lines), and it is therefore too weak to sustain a contracted drop. 

\emph{Autophobing.}---Another mechanism must be responsible for the large $\theta$ measured for large $\phi$. We recall the surfactant-like nature of \mbox{1,2-HD} molecules. Surfactant-containing liquids have long been known to autophobe, a phenomenon where $\theta$ increases due to modification of the solid surface energy by the surfactant molecules \cite{Rupprecht1991, Scales1986, Birch1994, Birch1995}. To the best of our knowledge, autophobing has not been reported or observed for mixtures of water and vicinal alkane diols, and never in combination with Marangoni contraction. 

\begin{figure}[t] 
	\begin{center}
		\includegraphics{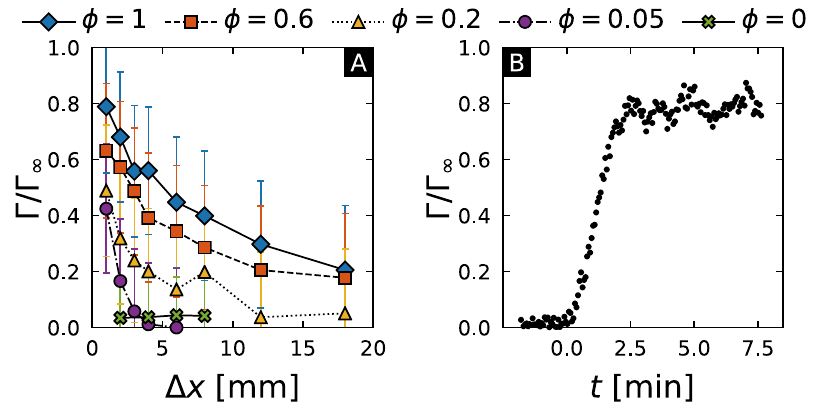} 
		\caption{(a) Normalized adsorption density ($\Gamma/\Gamma_\infty$) as a function of distance to the contact line ($\Delta x$) for several water--\mbox{1,2-HD} mixtures. (b) Temporal adsorption dynamics of pure \mbox{1,2-HD} at $\Delta x \approx 5$ mm. The liquid is deposited at $t~=~0$.}
		\label{fig3}
	\end{center}
\end{figure}

To induce autophobing, surfactant molecules have to adsorb on the solid-liquid interface (inside the drop) or on the solid-vapor interface (outside the drop), resulting in an overall decrease of $\gamma_\mathrm{SV}-\gamma_\mathrm{SL}$, where $\gamma_\mathrm{SV}$ is the surface tension of the solid-vapor interface, and $\gamma_\mathrm{SL}$ is the surface tension of the solid-liquid interface. In Fig.~\ref{fig3}(a) we report the adsorption properties of water--\mbox{1,2-HD} mixtures under ambient conditions, measured using ellipsometry (experimental details in Supplementary Material \cite{Note1}) \cite{Novotny1991}. Here, $\Gamma$ is the number density of adsorbed \mbox{1,2-HD} molecules normalized by $\Gamma_\infty$, the number density of adsorbed molecules corresponding to saturated coverage (measured in a closed chamber with saturated \mbox{1,2-HD} vapor). All values of $\Gamma/\Gamma_\infty$ were obtained after equilibrium was reached, as determined by measuring the temporal evolution of the adsorbed layer (Fig.~\ref{fig3}(b)). Equilibrium is typically reached within a few minutes after deposition of the liquid. The adsorbed layer is in equilibrium with the surrounding atmosphere (or vapor) -- \mbox{1,2-HD} molecules are continually adsorbing and desorbing \cite{Novotny1991, Zhu1991}. Complete desorption of the adsorption layer upon removal of the liquid typically takes an order of magnitude longer than the time it takes for the layer to adsorb (see Supplementary Material \cite{Note1}). 

Figure~\ref{fig3}(a) shows clear evidence of the adsorption of \mbox{1,2-HD} molecules on the substrate. Additionally, it shows that $\Gamma/\Gamma_\infty$ decreases both with the distance to the contact line $\Delta x$ and with $\phi$. This indicates that the concentration of \mbox{1,2-HD} in the vapor surrounding the drop is of key importance to the equilibrium surface concentration of molecules adsorbed on the substrate. As we increase $\Delta x$ or decrease $\phi$, the concentration of \mbox{1,2-HD} molecules in the vapor decreases. Hence, a lower amount of \mbox{1,2-HD} molecules is available in the vapor to adsorb on the substrate, while water becomes more abundant. Therefore, water coverage increases with increasing $\Delta x$ and decreasing $\phi$, resulting in a lower $\Gamma/\Gamma_\infty$. 

This indeed offers a direct explanation of the result in Fig.~\ref{fig1}, even when $\phi \gg \phi_\mathrm{CMC}$, where $\theta$ increases with $\phi$ and decreases with RH. An increase in RH leads to a lower $\Gamma/\Gamma_\infty$, due to the increased water coverage. Conversely, the \mbox{1,2-HD} coverage increases by increasing $\phi$. Adsorbed molecules change the surface energy of the substrate \cite{Bera2016}. The hydrophilic heads of the \mbox{1,2-HD} molecules attach to the substrate, and the hydrophobic tails are exposed to the ambient, thereby making the substrate more hydrophobic. This offers clear and direct evidence that the contact angles of autophobed drops depend on the RH of the close surrounding of the contact line. We remind that the internal flow is very weak at large $\phi$ (Fig.~\ref{fig2}(b)), for which we thus expect to recover the true equilibrium contact angle. In Young's law, which remains valid at equilibrium in the presence of surfactants \cite{Thiele2018}, the increased hydrophobicity of the substrate is reflected in the $\gamma_\mathrm{SV} - \gamma_\mathrm{SL}$ term, which becomes smaller with increasing $\Gamma/\Gamma_\infty$. Consequently, $\theta$ must increase, even though $\gamma_\mathrm{LV}$ remains constant above the CMC. 

Contrary to many previous works on autophobing \cite{Marmur1981, Frank1996, Takenaka2014, Bera2016, Zhong2016, Bera2018, Tadmor2019}, we do not see an initial spreading phase followed by a retraction to the quasi-steady $\theta$ (see Supplementary Material \cite{Note1}). This is likely due to the relatively high diffusion coefficient of \mbox{1,2-HD}, which is a result of  its small molecular size in comparison to other more common surfactants \cite{Kirkwood1948, Tanford1972}. The region of the substrate that is sampled by the liquid in determining the stationary $\theta$ is no larger than $10$ $\mu$m \cite{Decker1999}. The timescale associated with forming the equilibrium adsorption layer within this region is smaller than the spreading timescale \cite{Frank1995}, which is relatively long due to the high viscosity of \mbox{1,2-HD} ($\eta~\approx~82~$mPa$\cdot$s \cite{Jarosiewicz2004}). 

\emph{Effect of the chain length.}---Our experiments show that water--\mbox{1,2-HD} mixtures exhibit a transition between Marangoni contraction and autophobing. How generic is the observed competition between Marangoni contraction and authophobing for mixtures of water and vicinal alkane diols? Here, we address this question by considering three shorter vicinal alkane diols: 1,2-propanediol (\mbox{1,2-PrD}), 1,2-butanediol (\mbox{1,2-BD}), and 1,2-pentanediol (\mbox{1,2-PeD}), which have three, four, and five carbon atoms in their chain, respectively. These molecules allow a systematic study of the influence of the aliphatic chain length on the wetting properties. All diols are non-volatile and have a low $\gamma_\mathrm{LV}$ \cite{Karpitschka2010}. Short chain alkane diols show weaker surfactant-like behavior due to the decreased hydrophobicity of the molecule \cite{Smit1990}. 

\begin{figure}[t] 
	\begin{center}          
		\includegraphics{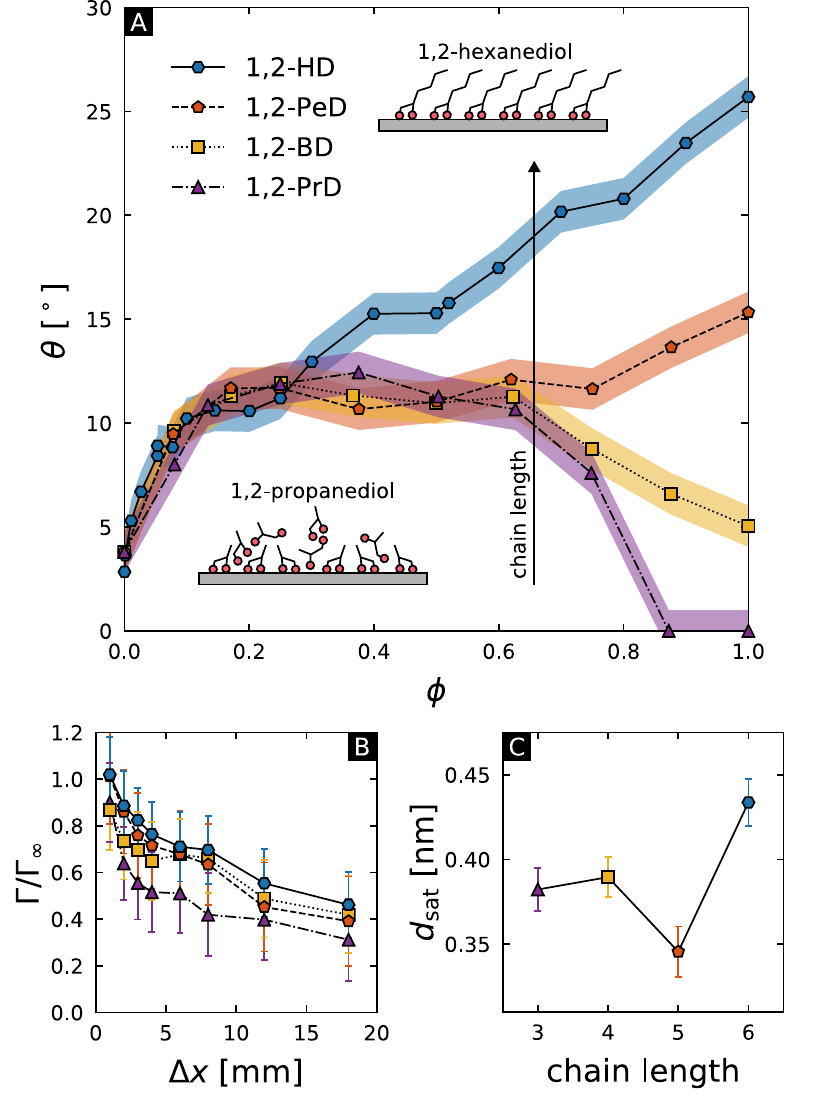} 
		\caption{(a) Contact angle ($\theta$) as a function of mass fraction ($\phi$) for several mixtures of water and vicinal alkane diols (RH = 60$\%$). The schematics show the structure of adsorbed 1,2-propanediol molecules and 1,2-hexanediol molecules.
 (b) Normalized adsorption density ($\Gamma/\Gamma_\infty$) as a function of distance to the contact line ($\Delta x$). (c) Thickness of the saturated film ($d_\mathrm{sat}$) for several vicinal alkane diols.}
		\label{fig4}
	\end{center}
\end{figure}

We study the properties of these diols using the same procedure as we used for \mbox{1,2-HD}. Figure \ref{fig4}(a) shows $\theta$ as a function of $\phi$ at RH~$\approx$~60\%. Starting at small $\phi$, we see that all diols follow a universal curve. This is perfectly consistent with Marangoni contraction, since this hydrodynamic mechanism is expected to be insensitive to molecular details. By contrast, the curves start to diverge and the length of the aliphatic chain matters for larger $\phi$ -- consistent with autophobing. The longest diol studied in this Letter, \mbox{1,2-HD}, exhibits strong autophobing behavior. As we move to short chain diols, the autophobing strength becomes smaller, indicated by smaller values of $\theta$ at $\phi = 1$. Additionally, Fig.~\ref{fig4}(a) shows that  Marangoni contraction dominates over autophobing up to a larger $\phi$ for shorter diols. While for \mbox{1,2-HD}, autophobing dominates of Marangoni contraction starting from $\phi \approx 0.3$, for \mbox{1,2-PrD}, by contrast, the full range of $\phi$ is consistent with Marangoni contraction -- there is no autophobing at all.

All four molecules adsorb on the substrate, as seen from the ellipsometry measurements presented in Fig.~\ref{fig4}(b). However, the reduced autophobing strength of the shorter diols is caused by the shorter hydrophobic chain in these molecules. The distance between the hydrophilic and hydrophobic parts of the molecule is smaller in shorter chain molecules, meaning that the polar nature of the hydroxyl groups becomes more relevant for the surface energy of an adsorbed layer of a short chain molecule such as \mbox{1,2-PrD}. The result is a more hydrophilic surface and therefore a smaller $\theta$. Figure~\ref{fig4}(b) shows that all diols studied here adsorb onto the substrate with similar $\Gamma/\Gamma_\infty$. However, as shown in Fig.~\ref{fig4}(c), not all adsorb in the same way as \mbox{1,2-HD}. Despite their small size, the saturated thickness $d_\mathrm{sat}$ of \mbox{1,2-PrD} and \mbox{1,2-BD} is larger than that of \mbox{1,2-PeD}, and only slightly smaller than that of \mbox{1,2-HD}. This indicates that \mbox{1,2-PrD} and \mbox{1,2-BD} do not form perfect monolayers, but form multi-layer disordered structures on the substrate \cite{Hu1995}. This is possible due to their relatively short aliphatic chain length, which leaves their hydroxyl groups partially exposed after one monolayer is adsorbed and allows for the adsorption of another (disordered) layer, similar to multi-layered structures of adsorbed water molecules \cite{Hu1995}. Note that the second layer will not be as strongly bound as the first layer that is directly adsorbed onto the surface and thus it cannot be considered an ordered, multilayer system (as seen in the schematic for \mbox{1,2-PrD} in Fig.~\ref{fig4}(a)). Hence, the surface energy of the substrate is not strongly affected by the adsorption of \mbox{1,2-PrD} and \mbox{1,2-BD} molecules. By contrast, \mbox{1,2-PeD} and \mbox{1,2-HD} adsorb in a monolayer structure (see the schematic for \mbox{1,2-HD} in Fig.~\ref{fig4}(a)), indicated by the increasing thickness between \mbox{1,2-PeD} and \mbox{1,2-HD} in Fig.~\ref{fig4}(c), and the fact that the layer thickness of \mbox{1,2-PeD} is lower than that of \mbox{1,2-PrD} and \mbox{1,2-BD}. This means that the long aliphatic chains are exposed --which increases the hydrophobicity of the surface-- and makes the formation of hydrogen bonds between the hydroxyl groups very unlikely. Therefore, autophobing occurs at large $\phi$ for molecules with a long aliphatic chain, due to the strong effect of the adsorbed molecules on the surface energy of the solid. By contrast, adsorbed molecules with a short aliphatic chain have little effect on the surface energy of the solid and Marangoni contraction dominates over the full range of $\phi$.

\emph{Conclusion}.---We have shown that the contact angles of two-component drops consisting of water and a vicinal alkane diol are determined by two separate mechanisms. A strong internal flow leads to an unexpectedly high $\theta$ for small $\phi$ drops, this mechanism is known as Marangoni contraction, and is of hydrodynamic nature. At large $\phi$, some mixtures of water and vicinal alkane diols autophobe due to the adsorption of diol molecules on the substrate. This changes the surface energy of the substrate leading to an increase of $\theta$. The autophobing strength depends on the length of the aliphatic chain of the diol. One can thus tune $\theta$ over a large range by selecting the correct diol (or surfactant) and a particular combination of $\phi$ and RH. Importantly, these mechanisms are generic and expected to be present in most mixtures containing (volatile) surfactant-like liquids.  

Precise control over the contact angle of a drop is important in many industrial and natural processes. Our result may be particularly interesting for applications that require high contact angles of drops consisting of low surface tension liquids, such as inkjet printing \cite{Wijshoff2018} or semiconductor processing \cite{Leenaars1990}. Further efforts should be devoted to a comprehensive theoretical description of the wetting properties in the autophobing regime, for example following the method of Thiele et al. \cite{Thiele2018}.

M. A. H., W. K. and O. R.-S. contributed equally to this manuscript. We thank M. Flapper, W. Tewes and H. Wijshoff for stimulating discussions. Financial support from an Industrial Partnership Programme of the Netherlands Organisation for Scientific Research (NWO), co-financed by Canon Production Printing, University of Twente, and Eindhoven University of Technology, and from the University of Twente--Max Planck Center ``Complex fluid dynamics--Fluid dynamics of Complexity'' is acknowledged.

\nocite{*}

\providecommand{\noopsort}[1]{}\providecommand{\singleletter}[1]{#1}%


\begin{thebibliography}{50}%
\makeatletter
\providecommand \@ifxundefined [1]{%
 \@ifx{#1\undefined}
}%
\providecommand \@ifnum [1]{%
 \ifnum #1\expandafter \@firstoftwo
 \else \expandafter \@secondoftwo
 \fi
}%
\providecommand \@ifx [1]{%
 \ifx #1\expandafter \@firstoftwo
 \else \expandafter \@secondoftwo
 \fi
}%
\providecommand \natexlab [1]{#1}%
\providecommand \enquote  [1]{``#1''}%
\providecommand \bibnamefont  [1]{#1}%
\providecommand \bibfnamefont [1]{#1}%
\providecommand \citenamefont [1]{#1}%
\providecommand \href@noop [0]{\@secondoftwo}%
\providecommand \href [0]{\begingroup \@sanitize@url \@href}%
\providecommand \@href[1]{\@@startlink{#1}\@@href}%
\providecommand \@@href[1]{\endgroup#1\@@endlink}%
\providecommand \@sanitize@url [0]{\catcode `\\12\catcode `\$12\catcode
  `\&12\catcode `\#12\catcode `\^12\catcode `\_12\catcode `\%12\relax}%
\providecommand \@@startlink[1]{}%
\providecommand \@@endlink[0]{}%
\providecommand \url  [0]{\begingroup\@sanitize@url \@url }%
\providecommand \@url [1]{\endgroup\@href {#1}{\urlprefix }}%
\providecommand \urlprefix  [0]{URL }%
\providecommand \Eprint [0]{\href }%
\providecommand \doibase [0]{https://doi.org/}%
\providecommand \selectlanguage [0]{\@gobble}%
\providecommand \bibinfo  [0]{\@secondoftwo}%
\providecommand \bibfield  [0]{\@secondoftwo}%
\providecommand \translation [1]{[#1]}%
\providecommand \BibitemOpen [0]{}%
\providecommand \bibitemStop [0]{}%
\providecommand \bibitemNoStop [0]{.\EOS\space}%
\providecommand \EOS [0]{\spacefactor3000\relax}%
\providecommand \BibitemShut  [1]{\csname bibitem#1\endcsname}%
\let\auto@bib@innerbib\@empty
\bibitem [{\citenamefont {Starov}\ and\ \citenamefont
  {Butt}(2018)}]{Starov2018}%
  \BibitemOpen
  \bibfield  {author} {\bibinfo {author} {\bibfnamefont {V.~M.}\ \bibnamefont
  {Starov}}\ and\ \bibinfo {author} {\bibfnamefont {H.-J.}\ \bibnamefont
  {Butt}},\ }\bibfield  {title} {\bibinfo {title} {Editorial overview:
  Worldwide increasing interest in wetting phenomena},\ }\href
  {https://doi.org/10.1016/j.cocis.2018.06.005} {\bibfield  {journal} {\bibinfo
   {journal} {Curr.~Opin.~Colloid~Interface~Sci.}\ }\textbf {\bibinfo {volume}
  {36}},\ \bibinfo {pages} {A1} (\bibinfo {year} {2018})}\BibitemShut {NoStop}%
\bibitem [{\citenamefont {Wijshoff}(2010)}]{Wijshoff2010}%
  \BibitemOpen
  \bibfield  {author} {\bibinfo {author} {\bibfnamefont {H.}~\bibnamefont
  {Wijshoff}},\ }\bibfield  {title} {\bibinfo {title} {The dynamics of the
  piezo inkjet printhead operation},\ }\href
  {https://doi.org/10.1016/j.physrep.2010.03.003} {\bibfield  {journal}
  {\bibinfo  {journal} {Phys. Rep.}\ }\textbf {\bibinfo {volume} {491}},\
  \bibinfo {pages} {77} (\bibinfo {year} {2010})}\BibitemShut {NoStop}%
\bibitem [{\citenamefont {Haagh}\ \emph {et~al.}(2019)\citenamefont {Haagh},
  \citenamefont {Schilderink}, \citenamefont {Mugele},\ and\ \citenamefont
  {Duits}}]{Haagh2019}%
  \BibitemOpen
  \bibfield  {author} {\bibinfo {author} {\bibfnamefont {M.~E.~J.}\
  \bibnamefont {Haagh}}, \bibinfo {author} {\bibfnamefont {N.}~\bibnamefont
  {Schilderink}}, \bibinfo {author} {\bibfnamefont {F.}~\bibnamefont
  {Mugele}},\ and\ \bibinfo {author} {\bibfnamefont {M.~H.~G.}\ \bibnamefont
  {Duits}},\ }\bibfield  {title} {\bibinfo {title} {Wetting of mineral surfaces
  by fatty-acid-laden oil and brine: carbonate effect at elevated
  temperature},\ }\href {https://doi.org/10.1021/acs.energyfuels.9b01351}
  {\bibfield  {journal} {\bibinfo  {journal} {Energy Fuels}\ }\textbf {\bibinfo
  {volume} {33}},\ \bibinfo {pages} {9446} (\bibinfo {year}
  {2019})}\BibitemShut {NoStop}%
\bibitem [{\citenamefont {Winkels}\ \emph {et~al.}(2011)\citenamefont
  {Winkels}, \citenamefont {Peters}, \citenamefont {Evangelista}, \citenamefont
  {Riepen}, \citenamefont {Daerr}, \citenamefont {Limat},\ and\ \citenamefont
  {Snoeijer}}]{Winkels2011}%
  \BibitemOpen
  \bibfield  {author} {\bibinfo {author} {\bibfnamefont {K.}~\bibnamefont
  {Winkels}}, \bibinfo {author} {\bibfnamefont {I.}~\bibnamefont {Peters}},
  \bibinfo {author} {\bibfnamefont {F.}~\bibnamefont {Evangelista}}, \bibinfo
  {author} {\bibfnamefont {M.}~\bibnamefont {Riepen}}, \bibinfo {author}
  {\bibfnamefont {A.}~\bibnamefont {Daerr}}, \bibinfo {author} {\bibfnamefont
  {L.}~\bibnamefont {Limat}},\ and\ \bibinfo {author} {\bibfnamefont
  {J.}~\bibnamefont {Snoeijer}},\ }\bibfield  {title} {\bibinfo {title}
  {Receding contact lines: from sliding drops to immersion lithography},\
  }\href {https://doi.org/10.1140/epjst/e2011-01374-6} {\bibfield  {journal}
  {\bibinfo  {journal} {Eur.~Phys.~J.~Special Topics}\ }\textbf {\bibinfo
  {volume} {192}},\ \bibinfo {pages} {195} (\bibinfo {year}
  {2011})}\BibitemShut {NoStop}%
\bibitem [{\citenamefont {Young}(1805)}]{Young1805}%
  \BibitemOpen
  \bibfield  {author} {\bibinfo {author} {\bibfnamefont {T.}~\bibnamefont
  {Young}},\ }\bibfield  {title} {\bibinfo {title} {An essay on the cohesion of
  fluids},\ }\href@noop {} {\bibfield  {journal} {\bibinfo  {journal}
  {Philos.~Trans.~Soc.~London}\ }\textbf {\bibinfo {volume} {95}},\ \bibinfo
  {pages} {65} (\bibinfo {year} {1805})}\BibitemShut {NoStop}%
\bibitem [{\citenamefont {Bonn}\ \emph {et~al.}(2009)\citenamefont {Bonn},
  \citenamefont {Eggers}, \citenamefont {Indekeu}, \citenamefont {Meunier},\
  and\ \citenamefont {Rolley}}]{Bonn2009}%
  \BibitemOpen
  \bibfield  {author} {\bibinfo {author} {\bibfnamefont {D.}~\bibnamefont
  {Bonn}}, \bibinfo {author} {\bibfnamefont {J.}~\bibnamefont {Eggers}},
  \bibinfo {author} {\bibfnamefont {J.}~\bibnamefont {Indekeu}}, \bibinfo
  {author} {\bibfnamefont {J.}~\bibnamefont {Meunier}},\ and\ \bibinfo {author}
  {\bibfnamefont {E.}~\bibnamefont {Rolley}},\ }\bibfield  {title} {\bibinfo
  {title} {Wetting and spreading},\ }\href
  {https://doi.org/10.1103/RevModPhys.81.739} {\bibfield  {journal} {\bibinfo
  {journal} {Rev.~Mod.~Phys.}\ }\textbf {\bibinfo {volume} {81}},\ \bibinfo
  {pages} {739} (\bibinfo {year} {2009})}\BibitemShut {NoStop}%
\bibitem [{\citenamefont {de~Gennes}\ \emph {et~al.}(2004)\citenamefont
  {de~Gennes}, \citenamefont {Brochard-Wyart},\ and\ \citenamefont
  {Qu{\'e}r{\'e}}}]{deGennes2004}%
  \BibitemOpen
  \bibfield  {author} {\bibinfo {author} {\bibfnamefont {P.-G.}\ \bibnamefont
  {de~Gennes}}, \bibinfo {author} {\bibfnamefont {F.}~\bibnamefont
  {Brochard-Wyart}},\ and\ \bibinfo {author} {\bibfnamefont {D.}~\bibnamefont
  {Qu{\'e}r{\'e}}},\ }\href {https://doi.org/10.1007/978-0-387-21656-0} {\emph
  {\bibinfo {title} {Capillarity and Wetting Phenomena: Drops, Bubbles, Pearls,
  Waves}}}\ (\bibinfo  {publisher} {Springer, New York},\ \bibinfo {year}
  {2004})\BibitemShut {NoStop}%
\bibitem [{\citenamefont {Drelich}\ \emph {et~al.}(2020)\citenamefont
  {Drelich}, \citenamefont {Boinovich}, \citenamefont {Chibowski},
  \citenamefont {Volpe}, \citenamefont {Ho{\l}ysz}, \citenamefont {Marmur},\
  and\ \citenamefont {Siboni}}]{Drelich2020}%
  \BibitemOpen
  \bibfield  {author} {\bibinfo {author} {\bibfnamefont {J.~W.}\ \bibnamefont
  {Drelich}}, \bibinfo {author} {\bibfnamefont {L.}~\bibnamefont {Boinovich}},
  \bibinfo {author} {\bibfnamefont {E.}~\bibnamefont {Chibowski}}, \bibinfo
  {author} {\bibfnamefont {C.~D.}\ \bibnamefont {Volpe}}, \bibinfo {author}
  {\bibfnamefont {L.}~\bibnamefont {Ho{\l}ysz}}, \bibinfo {author}
  {\bibfnamefont {A.}~\bibnamefont {Marmur}},\ and\ \bibinfo {author}
  {\bibfnamefont {S.}~\bibnamefont {Siboni}},\ }\bibfield  {title} {\bibinfo
  {title} {Contact angles: history of over 200 years of open questions},\
  }\href {https://doi.org/10.1680/jsuin.19.00007} {\bibfield  {journal}
  {\bibinfo  {journal} {Surface Innovations}\ }\textbf {\bibinfo {volume}
  {8}},\ \bibinfo {pages} {3} (\bibinfo {year} {2020})}\BibitemShut {NoStop}%
\bibitem [{\citenamefont {Leenaars}\ \emph {et~al.}(1990)\citenamefont
  {Leenaars}, \citenamefont {Huethorst},\ and\ \citenamefont {van
  Oekel}}]{Leenaars1990}%
  \BibitemOpen
  \bibfield  {author} {\bibinfo {author} {\bibfnamefont {A.~F.~M.}\
  \bibnamefont {Leenaars}}, \bibinfo {author} {\bibfnamefont {J.~A.~M.}\
  \bibnamefont {Huethorst}},\ and\ \bibinfo {author} {\bibfnamefont {J.~J.}\
  \bibnamefont {van Oekel}},\ }\bibfield  {title} {\bibinfo {title} {Marangoni
  drying: a new extremely clean drying process},\ }\href
  {https://doi.org/10.1021/la00101a014} {\bibfield  {journal} {\bibinfo
  {journal} {Langmuir}\ }\textbf {\bibinfo {volume} {6}},\ \bibinfo {pages}
  {1701} (\bibinfo {year} {1990})}\BibitemShut {NoStop}%
\bibitem [{\citenamefont {Matar}\ and\ \citenamefont
  {Craster}(2009)}]{Matar2016}%
  \BibitemOpen
  \bibfield  {author} {\bibinfo {author} {\bibfnamefont {O.~K.}\ \bibnamefont
  {Matar}}\ and\ \bibinfo {author} {\bibfnamefont {R.~V.}\ \bibnamefont
  {Craster}},\ }\bibfield  {title} {\bibinfo {title} {Dynamics of
  surfactant-assisted spreading},\ }\href {https://doi.org/10.1039/b908719m}
  {\bibfield  {journal} {\bibinfo  {journal} {Soft Matter}\ }\textbf {\bibinfo
  {volume} {5}},\ \bibinfo {pages} {3801} (\bibinfo {year} {2009})}\BibitemShut
  {NoStop}%
\bibitem [{\citenamefont {Tan}\ \emph {et~al.}(2016)\citenamefont {Tan},
  \citenamefont {Diddens}, \citenamefont {Lv}, \citenamefont {Kuerten},
  \citenamefont {Zhang},\ and\ \citenamefont {Lohse}}]{Tan2016}%
  \BibitemOpen
  \bibfield  {author} {\bibinfo {author} {\bibfnamefont {H.}~\bibnamefont
  {Tan}}, \bibinfo {author} {\bibfnamefont {C.}~\bibnamefont {Diddens}},
  \bibinfo {author} {\bibfnamefont {P.}~\bibnamefont {Lv}}, \bibinfo {author}
  {\bibfnamefont {J.~G.~M.}\ \bibnamefont {Kuerten}}, \bibinfo {author}
  {\bibfnamefont {X.}~\bibnamefont {Zhang}},\ and\ \bibinfo {author}
  {\bibfnamefont {D.}~\bibnamefont {Lohse}},\ }\bibfield  {title} {\bibinfo
  {title} {Evaporation-triggered microdroplet nucleation and the four life
  phases of an evaaporating ouzo drop},\ }\href
  {https://doi.org/10.1073/pnas.1602260113} {\bibfield  {journal} {\bibinfo
  {journal} {Proc.~Natl.~Acad.~Sci.~U.S.A}\ }\textbf {\bibinfo {volume}
  {113}},\ \bibinfo {pages} {8642} (\bibinfo {year} {2016})}\BibitemShut
  {NoStop}%
\bibitem [{\citenamefont {Li}\ \emph {et~al.}(2018)\citenamefont {Li},
  \citenamefont {Lv}, \citenamefont {Diddens}, \citenamefont {Tan},
  \citenamefont {Wijshoff}, \citenamefont {Versluis},\ and\ \citenamefont
  {Lohse}}]{Li2018}%
  \BibitemOpen
  \bibfield  {author} {\bibinfo {author} {\bibfnamefont {Y.}~\bibnamefont
  {Li}}, \bibinfo {author} {\bibfnamefont {P.}~\bibnamefont {Lv}}, \bibinfo
  {author} {\bibfnamefont {C.}~\bibnamefont {Diddens}}, \bibinfo {author}
  {\bibfnamefont {H.}~\bibnamefont {Tan}}, \bibinfo {author} {\bibfnamefont
  {H.}~\bibnamefont {Wijshoff}}, \bibinfo {author} {\bibfnamefont
  {M.}~\bibnamefont {Versluis}},\ and\ \bibinfo {author} {\bibfnamefont
  {D.}~\bibnamefont {Lohse}},\ }\bibfield  {title} {\bibinfo {title}
  {Evaporation-triggered segregation of sessile binary droplets},\ }\href
  {https://doi.org/10.1103/PhysRevLett.120.224501} {\bibfield  {journal}
  {\bibinfo  {journal} {Phys.~Rev.~Lett.}\ }\textbf {\bibinfo {volume} {120}},\
  \bibinfo {pages} {224501} (\bibinfo {year} {2018})}\BibitemShut {NoStop}%
\bibitem [{\citenamefont {Sefiane}\ \emph {et~al.}(2008)\citenamefont
  {Sefiane}, \citenamefont {David},\ and\ \citenamefont
  {Shanahan}}]{Sefiane2008}%
  \BibitemOpen
  \bibfield  {author} {\bibinfo {author} {\bibfnamefont {K.}~\bibnamefont
  {Sefiane}}, \bibinfo {author} {\bibfnamefont {S.}~\bibnamefont {David}},\
  and\ \bibinfo {author} {\bibfnamefont {M.~E.~R.}\ \bibnamefont {Shanahan}},\
  }\bibfield  {title} {\bibinfo {title} {Wetting and evaporation of binary
  mixture drops},\ }\href {https://doi.org/10.1021/jp8030418} {\bibfield
  {journal} {\bibinfo  {journal} {J.~Phys.~Chem.~B}\ }\textbf {\bibinfo
  {volume} {112}},\ \bibinfo {pages} {11317} (\bibinfo {year}
  {2008})}\BibitemShut {NoStop}%
\bibitem [{\citenamefont {Keiser}\ \emph {et~al.}(2017)\citenamefont {Keiser},
  \citenamefont {Bense}, \citenamefont {Colinet}, \citenamefont {Bico},\ and\
  \citenamefont {Reyssat}}]{Keiser2017}%
  \BibitemOpen
  \bibfield  {author} {\bibinfo {author} {\bibfnamefont {L.}~\bibnamefont
  {Keiser}}, \bibinfo {author} {\bibfnamefont {H.}~\bibnamefont {Bense}},
  \bibinfo {author} {\bibfnamefont {P.}~\bibnamefont {Colinet}}, \bibinfo
  {author} {\bibfnamefont {J.}~\bibnamefont {Bico}},\ and\ \bibinfo {author}
  {\bibfnamefont {E.}~\bibnamefont {Reyssat}},\ }\bibfield  {title} {\bibinfo
  {title} {Marangoni bursting: Evaporation-induced emulsification of binary
  mixtures on a liquid layer},\ }\href
  {https://doi.org/10.1103/PhysRevLett.118.074504} {\bibfield  {journal}
  {\bibinfo  {journal} {Phys.~Rev.~Lett.}\ }\textbf {\bibinfo {volume} {118}},\
  \bibinfo {pages} {074504} (\bibinfo {year} {2017})}\BibitemShut {NoStop}%
\bibitem [{\citenamefont {Kim}\ \emph {et~al.}(2016)\citenamefont {Kim},
  \citenamefont {Boulogne}, \citenamefont {Um}, \citenamefont {Jacobi},
  \citenamefont {Button},\ and\ \citenamefont {Stone}}]{Kim2016}%
  \BibitemOpen
  \bibfield  {author} {\bibinfo {author} {\bibfnamefont {H.}~\bibnamefont
  {Kim}}, \bibinfo {author} {\bibfnamefont {F.}~\bibnamefont {Boulogne}},
  \bibinfo {author} {\bibfnamefont {E.}~\bibnamefont {Um}}, \bibinfo {author}
  {\bibfnamefont {I.}~\bibnamefont {Jacobi}}, \bibinfo {author} {\bibfnamefont
  {E.}~\bibnamefont {Button}},\ and\ \bibinfo {author} {\bibfnamefont {H.~A.}\
  \bibnamefont {Stone}},\ }\bibfield  {title} {\bibinfo {title} {Controlled
  uniform coating from the interplay of marangoni flows and surface-adsorbed
  macromolecules},\ }\href {https://doi.org/10.1103/PhysRevLett.116.124501}
  {\bibfield  {journal} {\bibinfo  {journal} {Phys.~Rev.~Lett.}\ }\textbf
  {\bibinfo {volume} {116}},\ \bibinfo {pages} {124501} (\bibinfo {year}
  {2016})}\BibitemShut {NoStop}%
\bibitem [{Note1()}]{Note1}%
  \BibitemOpen
  \bibinfo {note} {See Supplemental Material at WEBSITE LINK\ for additional
  information on experimental details and a figure showing the attempted
  collapse the contact angle data using the Marangoni contraction scaling
  law.}\BibitemShut {Stop}%
\bibitem [{\citenamefont {Hajji}\ \emph {et~al.}(1989)\citenamefont {Hajji},
  \citenamefont {Errahmani}, \citenamefont {Coudert}, \citenamefont {Durand},
  \citenamefont {Cao},\ and\ \citenamefont {Taillandier}}]{Haiji1989}%
  \BibitemOpen
  \bibfield  {author} {\bibinfo {author} {\bibfnamefont {S.}~\bibnamefont
  {Hajji}}, \bibinfo {author} {\bibfnamefont {M.}~\bibnamefont {Errahmani}},
  \bibinfo {author} {\bibfnamefont {R.}~\bibnamefont {Coudert}}, \bibinfo
  {author} {\bibfnamefont {R.}~\bibnamefont {Durand}}, \bibinfo {author}
  {\bibfnamefont {A.}~\bibnamefont {Cao}},\ and\ \bibinfo {author}
  {\bibfnamefont {E.}~\bibnamefont {Taillandier}},\ }\bibfield  {title}
  {\bibinfo {title} {A comparative study of hexanediol-1,2 and
  octanetriol-1,2,3 in aqueous solutions by different physical techniques},\
  }\href {https://doi.org/10.1021/j100349a028} {\bibfield  {journal} {\bibinfo
  {journal} {J.~Phys.~Chem.}\ }\textbf {\bibinfo {volume} {93}},\ \bibinfo
  {pages} {4819} (\bibinfo {year} {1989})}\BibitemShut {NoStop}%
\bibitem [{\citenamefont {Frindi}\ \emph {et~al.}(1991)\citenamefont {Frindi},
  \citenamefont {Michels},\ and\ \citenamefont {Zana}}]{Frindi1991}%
  \BibitemOpen
  \bibfield  {author} {\bibinfo {author} {\bibfnamefont {M.}~\bibnamefont
  {Frindi}}, \bibinfo {author} {\bibfnamefont {B.}~\bibnamefont {Michels}},\
  and\ \bibinfo {author} {\bibfnamefont {R.}~\bibnamefont {Zana}},\ }\bibfield
  {title} {\bibinfo {title} {Ultrasonic absorption studies of surfactant
  exchange between micelles and bulk phase in aqueous micellar solutions of
  nonionic surfactants with short alkyl chains 1. 1,2-hexanediol and
  1,2,3-octanetriol},\ }\href {https://doi.org/10.1021/j100165a044} {\bibfield
  {journal} {\bibinfo  {journal} {J.~Phys.~Chem.}\ }\textbf {\bibinfo {volume}
  {95}},\ \bibinfo {pages} {4832} (\bibinfo {year} {1991})}\BibitemShut
  {NoStop}%
\bibitem [{\citenamefont {Sz{\'{e}}kely}\ \emph {et~al.}(2007)\citenamefont
  {Sz{\'{e}}kely}, \citenamefont {Alm{\'{a}}sy}, \citenamefont
  {R{\u{a}}dulescu},\ and\ \citenamefont {Rosta}}]{Szekely2007}%
  \BibitemOpen
  \bibfield  {author} {\bibinfo {author} {\bibfnamefont {N.}~\bibnamefont
  {Sz{\'{e}}kely}}, \bibinfo {author} {\bibfnamefont {L.}~\bibnamefont
  {Alm{\'{a}}sy}}, \bibinfo {author} {\bibfnamefont {A.}~\bibnamefont
  {R{\u{a}}dulescu}},\ and\ \bibinfo {author} {\bibfnamefont {L.}~\bibnamefont
  {Rosta}},\ }\bibfield  {title} {\bibinfo {title} {Small-angle neutron
  scattering study of aqueous solutions of pentanediol and hexanediol},\ }\href
  {https://doi.org/10.1107/S0021889807001483} {\bibfield  {journal} {\bibinfo
  {journal} {J.~Appl.~Cryst.}\ }\textbf {\bibinfo {volume} {40}},\ \bibinfo
  {pages} {s307} (\bibinfo {year} {2007})}\BibitemShut {NoStop}%
\bibitem [{\citenamefont {Tan}\ and\ \citenamefont {Finch}(2018)}]{Tan2018}%
  \BibitemOpen
  \bibfield  {author} {\bibinfo {author} {\bibfnamefont {Y.~H.}\ \bibnamefont
  {Tan}}\ and\ \bibinfo {author} {\bibfnamefont {J.~A.}\ \bibnamefont
  {Finch}},\ }\bibfield  {title} {\bibinfo {title} {Frothers and gas
  dispersion: a review of the structure-property function relationship},\
  }\href {https://doi.org/10.5277/ppmp1814} {\bibfield  {journal} {\bibinfo
  {journal} {Physicochem.~Probl.~Miner.~Process.}\ }\textbf {\bibinfo {volume}
  {54}},\ \bibinfo {pages} {40} (\bibinfo {year} {2018})}\BibitemShut {NoStop}%
\bibitem [{\citenamefont {Romero}\ \emph {et~al.}(2007)\citenamefont {Romero},
  \citenamefont {P{\'{a}}ez}, \citenamefont {Miranda}, \citenamefont
  {Hern{\'{a}}ndez},\ and\ \citenamefont {Oviedo}}]{Romero2007}%
  \BibitemOpen
  \bibfield  {author} {\bibinfo {author} {\bibfnamefont {C.~M.}\ \bibnamefont
  {Romero}}, \bibinfo {author} {\bibfnamefont {M.~S.}\ \bibnamefont
  {P{\'{a}}ez}}, \bibinfo {author} {\bibfnamefont {J.~A.}\ \bibnamefont
  {Miranda}}, \bibinfo {author} {\bibfnamefont {D.~J.}\ \bibnamefont
  {Hern{\'{a}}ndez}},\ and\ \bibinfo {author} {\bibfnamefont {L.~E.}\
  \bibnamefont {Oviedo}},\ }\bibfield  {title} {\bibinfo {title} {Effect of
  temperature on the surface tension of diluted aqueous solutions of
  1,2-hexanediol, 1,5-hexanediol, 1,6-hexanediol and 2,5-hexanediol},\ }\href
  {https://doi.org/10.1016/j.fluid.2007.05.029} {\bibfield  {journal} {\bibinfo
   {journal} {Fluid~Ph.~Equilibria}\ }\textbf {\bibinfo {volume} {258}},\
  \bibinfo {pages} {67} (\bibinfo {year} {2007})}\BibitemShut {NoStop}%
\bibitem [{\citenamefont {Karpitschka}\ \emph {et~al.}(2017)\citenamefont
  {Karpitschka}, \citenamefont {Liebig},\ and\ \citenamefont
  {Riegler}}]{Karpitschka2017}%
  \BibitemOpen
  \bibfield  {author} {\bibinfo {author} {\bibfnamefont {S.}~\bibnamefont
  {Karpitschka}}, \bibinfo {author} {\bibfnamefont {F.}~\bibnamefont
  {Liebig}},\ and\ \bibinfo {author} {\bibfnamefont {H.}~\bibnamefont
  {Riegler}},\ }\bibfield  {title} {\bibinfo {title} {Marangoni contraction of
  evaporating sessile droplets of binary mixtures},\ }\href
  {https://doi.org/10.1021/acs.langmuir.7b00740} {\bibfield  {journal}
  {\bibinfo  {journal} {Langmuir}\ }\textbf {\bibinfo {volume} {33}},\ \bibinfo
  {pages} {4682} (\bibinfo {year} {2017})}\BibitemShut {NoStop}%
\bibitem [{\citenamefont {Marra}\ and\ \citenamefont
  {Huethorst}(1991)}]{Marra1991}%
  \BibitemOpen
  \bibfield  {author} {\bibinfo {author} {\bibfnamefont {J.}~\bibnamefont
  {Marra}}\ and\ \bibinfo {author} {\bibfnamefont {J.~A.~M.}\ \bibnamefont
  {Huethorst}},\ }\bibfield  {title} {\bibinfo {title} {Physical principles of
  marangoni drying},\ }\href {https://doi.org/10.1021/la00059a057} {\bibfield
  {journal} {\bibinfo  {journal} {Langmuir}\ }\textbf {\bibinfo {volume} {7}},\
  \bibinfo {pages} {2748} (\bibinfo {year} {1991})}\BibitemShut {NoStop}%
\bibitem [{\citenamefont {Cira}\ \emph {et~al.}(2015)\citenamefont {Cira},
  \citenamefont {Benusiglio},\ and\ \citenamefont {Prakash}}]{Cira2015}%
  \BibitemOpen
  \bibfield  {author} {\bibinfo {author} {\bibfnamefont {N.~J.}\ \bibnamefont
  {Cira}}, \bibinfo {author} {\bibfnamefont {A.}~\bibnamefont {Benusiglio}},\
  and\ \bibinfo {author} {\bibfnamefont {M.}~\bibnamefont {Prakash}},\
  }\bibfield  {title} {\bibinfo {title} {Vapour-mediated sensing and motility
  in two-component droplets},\ }\href {https://doi.org/10.1038/nature14272}
  {\bibfield  {journal} {\bibinfo  {journal} {Nature}\ }\textbf {\bibinfo
  {volume} {519}},\ \bibinfo {pages} {446} (\bibinfo {year}
  {2015})}\BibitemShut {NoStop}%
\bibitem [{\citenamefont {Benusiglio}\ \emph {et~al.}(2018)\citenamefont
  {Benusiglio}, \citenamefont {Cira},\ and\ \citenamefont
  {Prakash}}]{Benusiglio2018}%
  \BibitemOpen
  \bibfield  {author} {\bibinfo {author} {\bibfnamefont {A.}~\bibnamefont
  {Benusiglio}}, \bibinfo {author} {\bibfnamefont {N.~J.}\ \bibnamefont
  {Cira}},\ and\ \bibinfo {author} {\bibfnamefont {M.}~\bibnamefont
  {Prakash}},\ }\bibfield  {title} {\bibinfo {title} {Two-component
  marangoni-contracted droplets: friction and shape},\ }\href
  {https://doi.org/10.1039/c7sm02361h} {\bibfield  {journal} {\bibinfo
  {journal} {Soft Matter}\ }\textbf {\bibinfo {volume} {14}},\ \bibinfo {pages}
  {7724} (\bibinfo {year} {2018})}\BibitemShut {NoStop}%
\bibitem [{\citenamefont {Deegan}\ \emph {et~al.}(1997)\citenamefont {Deegan},
  \citenamefont {Baakajin}, \citenamefont {Dupont}, \citenamefont {Huber},
  \citenamefont {Nagel},\ and\ \citenamefont {Witten}}]{Deegan1997}%
  \BibitemOpen
  \bibfield  {author} {\bibinfo {author} {\bibfnamefont {R.}~\bibnamefont
  {Deegan}}, \bibinfo {author} {\bibfnamefont {O.}~\bibnamefont {Baakajin}},
  \bibinfo {author} {\bibfnamefont {T.}~\bibnamefont {Dupont}}, \bibinfo
  {author} {\bibfnamefont {G.}~\bibnamefont {Huber}}, \bibinfo {author}
  {\bibfnamefont {S.}~\bibnamefont {Nagel}},\ and\ \bibinfo {author}
  {\bibfnamefont {T.}~\bibnamefont {Witten}},\ }\bibfield  {title} {\bibinfo
  {title} {Capillary flow as the cause of ring stains from dried liquid
  drops},\ }\href {https://doi.org/10.1038/39827} {\bibfield  {journal}
  {\bibinfo  {journal} {Nature}\ }\textbf {\bibinfo {volume} {389}},\ \bibinfo
  {pages} {827} (\bibinfo {year} {1997})}\BibitemShut {NoStop}%
\bibitem [{\citenamefont {Semenov}\ \emph {et~al.}(2014)\citenamefont
  {Semenov}, \citenamefont {Trybala}, \citenamefont {Rubio}, \citenamefont
  {Kovalchuk}, \citenamefont {Starov},\ and\ \citenamefont
  {Velarde}}]{Semenov2014}%
  \BibitemOpen
  \bibfield  {author} {\bibinfo {author} {\bibfnamefont {S.}~\bibnamefont
  {Semenov}}, \bibinfo {author} {\bibfnamefont {A.}~\bibnamefont {Trybala}},
  \bibinfo {author} {\bibfnamefont {R.~G.}\ \bibnamefont {Rubio}}, \bibinfo
  {author} {\bibfnamefont {N.}~\bibnamefont {Kovalchuk}}, \bibinfo {author}
  {\bibfnamefont {V.}~\bibnamefont {Starov}},\ and\ \bibinfo {author}
  {\bibfnamefont {M.~G.}\ \bibnamefont {Velarde}},\ }\bibfield  {title}
  {\bibinfo {title} {Simultaneous spreading and evaporation: recent
  developments},\ }\href {https://doi.org/10.1016/j.cis.2013.08.006} {\bibfield
   {journal} {\bibinfo  {journal} {Adv.~Colloid~Interface~Sci.}\ }\textbf
  {\bibinfo {volume} {206}},\ \bibinfo {pages} {382} (\bibinfo {year}
  {2014})}\BibitemShut {NoStop}%
\bibitem [{\citenamefont {Rupprecht}\ and\ \citenamefont
  {Gu}(1991)}]{Rupprecht1991}%
  \BibitemOpen
  \bibfield  {author} {\bibinfo {author} {\bibfnamefont {H.}~\bibnamefont
  {Rupprecht}}\ and\ \bibinfo {author} {\bibfnamefont {T.}~\bibnamefont {Gu}},\
  }\bibfield  {title} {\bibinfo {title} {Structure of adsorption layers of
  ionic surfactants at the solid/liquid interface},\ }\href
  {https://doi.org/10.1007/BF00655889} {\bibfield  {journal} {\bibinfo
  {journal} {Colloid~Polym.~Sci.}\ }\textbf {\bibinfo {volume} {269}},\
  \bibinfo {pages} {506} (\bibinfo {year} {1991})}\BibitemShut {NoStop}%
\bibitem [{\citenamefont {Scales}\ \emph {et~al.}(1986)\citenamefont {Scales},
  \citenamefont {Grieser}, \citenamefont {Furlong},\ and\ \citenamefont
  {Healy}}]{Scales1986}%
  \BibitemOpen
  \bibfield  {author} {\bibinfo {author} {\bibfnamefont {P.~J.}\ \bibnamefont
  {Scales}}, \bibinfo {author} {\bibfnamefont {F.}~\bibnamefont {Grieser}},
  \bibinfo {author} {\bibfnamefont {D.~N.}\ \bibnamefont {Furlong}},\ and\
  \bibinfo {author} {\bibfnamefont {T.~W.}\ \bibnamefont {Healy}},\ }\bibfield
  {title} {\bibinfo {title} {Contact angle changes for hydrophobic and
  hydrophilic surfaces induced by nonionic surfactants},\ }\href
  {https://doi.org/10.1016/0166-6622(86)80082-8} {\bibfield  {journal}
  {\bibinfo  {journal} {Colloids~Surf.}\ }\textbf {\bibinfo {volume} {21}},\
  \bibinfo {pages} {55} (\bibinfo {year} {1986})}\BibitemShut {NoStop}%
\bibitem [{\citenamefont {Birch}\ \emph {et~al.}(1994)\citenamefont {Birch},
  \citenamefont {Knewtson}, \citenamefont {Garoff}, \citenamefont {Suter},\
  and\ \citenamefont {Satija}}]{Birch1994}%
  \BibitemOpen
  \bibfield  {author} {\bibinfo {author} {\bibfnamefont {W.~R.}\ \bibnamefont
  {Birch}}, \bibinfo {author} {\bibfnamefont {M.~A.}\ \bibnamefont {Knewtson}},
  \bibinfo {author} {\bibfnamefont {S.}~\bibnamefont {Garoff}}, \bibinfo
  {author} {\bibfnamefont {R.~M.}\ \bibnamefont {Suter}},\ and\ \bibinfo
  {author} {\bibfnamefont {S.}~\bibnamefont {Satija}},\ }\bibfield  {title}
  {\bibinfo {title} {The molecular stucture of autophobed monolayers and
  precursing films of a cationic surfactant on the silicon oxide/silicon
  surface},\ }\href {https://doi.org/10.1016/0927-7757(94)80114-2} {\bibfield
  {journal} {\bibinfo  {journal} {Colloids~Surf.~A}\ }\textbf {\bibinfo
  {volume} {89}},\ \bibinfo {pages} {145} (\bibinfo {year} {1994})}\BibitemShut
  {NoStop}%
\bibitem [{\citenamefont {Birch}\ \emph {et~al.}(1995)\citenamefont {Birch},
  \citenamefont {Knewtson}, \citenamefont {Garoff}, \citenamefont {Suter},\
  and\ \citenamefont {Satija}}]{Birch1995}%
  \BibitemOpen
  \bibfield  {author} {\bibinfo {author} {\bibfnamefont {W.~R.}\ \bibnamefont
  {Birch}}, \bibinfo {author} {\bibfnamefont {M.~A.}\ \bibnamefont {Knewtson}},
  \bibinfo {author} {\bibfnamefont {S.}~\bibnamefont {Garoff}}, \bibinfo
  {author} {\bibfnamefont {R.~M.}\ \bibnamefont {Suter}},\ and\ \bibinfo
  {author} {\bibfnamefont {S.}~\bibnamefont {Satija}},\ }\bibfield  {title}
  {\bibinfo {title} {Structure of precursing thin films of an anionic
  surfactant on a silicon oxide/silicon surface},\ }\href
  {https://doi.org/10.1021/la00001a012} {\bibfield  {journal} {\bibinfo
  {journal} {Langmuir}\ }\textbf {\bibinfo {volume} {11}},\ \bibinfo {pages}
  {48} (\bibinfo {year} {1995})}\BibitemShut {NoStop}%
\bibitem [{\citenamefont {Novotny}\ and\ \citenamefont
  {Marmur}(1991)}]{Novotny1991}%
  \BibitemOpen
  \bibfield  {author} {\bibinfo {author} {\bibfnamefont {V.~J.}\ \bibnamefont
  {Novotny}}\ and\ \bibinfo {author} {\bibfnamefont {A.}~\bibnamefont
  {Marmur}},\ }\bibfield  {title} {\bibinfo {title} {Wetting autophobicity},\
  }\href {https://doi.org/10.1016/0021-9797(91)90367-H} {\bibfield  {journal}
  {\bibinfo  {journal} {J.~Colloid~Interface~Sci.}\ }\textbf {\bibinfo {volume}
  {145}},\ \bibinfo {pages} {355} (\bibinfo {year} {1991})}\BibitemShut
  {NoStop}%
\bibitem [{\citenamefont {Zhu}\ and\ \citenamefont {Gu}(1991)}]{Zhu1991}%
  \BibitemOpen
  \bibfield  {author} {\bibinfo {author} {\bibfnamefont {B.-Y.}\ \bibnamefont
  {Zhu}}\ and\ \bibinfo {author} {\bibfnamefont {T.}~\bibnamefont {Gu}},\
  }\bibfield  {title} {\bibinfo {title} {Surfactant adsorption at solid-liquid
  interfaces},\ }\href {https://doi.org/10.1016/0001-8686(91)80037-K}
  {\bibfield  {journal} {\bibinfo  {journal} {Adv.~Colloid~Interface~Sci}\
  }\textbf {\bibinfo {volume} {37}},\ \bibinfo {pages} {1} (\bibinfo {year}
  {1991})}\BibitemShut {NoStop}%
\bibitem [{\citenamefont {Bera}\ \emph {et~al.}(2016)\citenamefont {Bera},
  \citenamefont {Duits}, \citenamefont {Stuart}, \citenamefont {van~den Ende},\
  and\ \citenamefont {Mugele}}]{Bera2016}%
  \BibitemOpen
  \bibfield  {author} {\bibinfo {author} {\bibfnamefont {B.}~\bibnamefont
  {Bera}}, \bibinfo {author} {\bibfnamefont {M.~H.~G.}\ \bibnamefont {Duits}},
  \bibinfo {author} {\bibfnamefont {M.~A.~C.}\ \bibnamefont {Stuart}}, \bibinfo
  {author} {\bibfnamefont {D.}~\bibnamefont {van~den Ende}},\ and\ \bibinfo
  {author} {\bibfnamefont {F.}~\bibnamefont {Mugele}},\ }\bibfield  {title}
  {\bibinfo {title} {Surfactant induced autophobing},\ }\href
  {https://doi.org/10.1039/C6SM00128A} {\bibfield  {journal} {\bibinfo
  {journal} {Soft Matter}\ }\textbf {\bibinfo {volume} {12}},\ \bibinfo {pages}
  {4562} (\bibinfo {year} {2016})}\BibitemShut {NoStop}%
\bibitem [{\citenamefont {Thiele}\ \emph {et~al.}(2018)\citenamefont {Thiele},
  \citenamefont {Snoeijer}, \citenamefont {Trinschek},\ and\ \citenamefont
  {John}}]{Thiele2018}%
  \BibitemOpen
  \bibfield  {author} {\bibinfo {author} {\bibfnamefont {U.}~\bibnamefont
  {Thiele}}, \bibinfo {author} {\bibfnamefont {J.~H.}\ \bibnamefont
  {Snoeijer}}, \bibinfo {author} {\bibfnamefont {S.}~\bibnamefont
  {Trinschek}},\ and\ \bibinfo {author} {\bibfnamefont {K.}~\bibnamefont
  {John}},\ }\bibfield  {title} {\bibinfo {title} {Equilibrium contact angle
  and adsorption layer properties with surfactants},\ }\href
  {https://doi.org/10.1021/acs.langmuir.8b00513} {\bibfield  {journal}
  {\bibinfo  {journal} {Langmuir}\ }\textbf {\bibinfo {volume} {34}},\ \bibinfo
  {pages} {7210} (\bibinfo {year} {2018})}\BibitemShut {NoStop}%
\bibitem [{\citenamefont {Marmur}\ and\ \citenamefont
  {Lelah}(1981)}]{Marmur1981}%
  \BibitemOpen
  \bibfield  {author} {\bibinfo {author} {\bibfnamefont {A.}~\bibnamefont
  {Marmur}}\ and\ \bibinfo {author} {\bibfnamefont {M.~D.}\ \bibnamefont
  {Lelah}},\ }\bibfield  {title} {\bibinfo {title} {The spreading of aqueous
  surfactant solutions on glass},\ }\href
  {https://doi.org/10.1080/00986448108910901} {\bibfield  {journal} {\bibinfo
  {journal} {Chem.~Eng.~Commun.}\ }\textbf {\bibinfo {volume} {13}},\ \bibinfo
  {pages} {133} (\bibinfo {year} {1981})}\BibitemShut {NoStop}%
\bibitem [{\citenamefont {Frank}\ and\ \citenamefont
  {Garoff}(1996)}]{Frank1996}%
  \BibitemOpen
  \bibfield  {author} {\bibinfo {author} {\bibfnamefont {B.}~\bibnamefont
  {Frank}}\ and\ \bibinfo {author} {\bibfnamefont {S.}~\bibnamefont {Garoff}},\
  }\bibfield  {title} {\bibinfo {title} {Surfactant self-assembly near contact
  lines: control of advancing surfactant solutions},\ }\href
  {https://doi.org/10.1016/0927-7757(96)03622-9} {\bibfield  {journal}
  {\bibinfo  {journal} {Colloids Surf. A}\ }\textbf {\bibinfo {volume} {116}},\
  \bibinfo {pages} {31} (\bibinfo {year} {1996})}\BibitemShut {NoStop}%
\bibitem [{\citenamefont {Takenaka}\ \emph {et~al.}(2014)\citenamefont
  {Takenaka}, \citenamefont {Sumino},\ and\ \citenamefont
  {Ohzono}}]{Takenaka2014}%
  \BibitemOpen
  \bibfield  {author} {\bibinfo {author} {\bibfnamefont {Y.}~\bibnamefont
  {Takenaka}}, \bibinfo {author} {\bibfnamefont {Y.}~\bibnamefont {Sumino}},\
  and\ \bibinfo {author} {\bibfnamefont {T.}~\bibnamefont {Ohzono}},\
  }\bibfield  {title} {\bibinfo {title} {Dewetting of a droplet induced by the
  adsorption of surfactants on a glass substrate},\ }\href
  {https://doi.org/10.1039/c4sm00798k} {\bibfield  {journal} {\bibinfo
  {journal} {Soft Matter}\ }\textbf {\bibinfo {volume} {10}},\ \bibinfo {pages}
  {5597} (\bibinfo {year} {2014})}\BibitemShut {NoStop}%
\bibitem [{\citenamefont {Zhong}\ and\ \citenamefont {Duan}(2016)}]{Zhong2016}%
  \BibitemOpen
  \bibfield  {author} {\bibinfo {author} {\bibfnamefont {X.}~\bibnamefont
  {Zhong}}\ and\ \bibinfo {author} {\bibfnamefont {F.}~\bibnamefont {Duan}},\
  }\bibfield  {title} {\bibinfo {title} {Dewetting transition induced by
  surfactants in sessile droplets at the early evaporation stage},\ }\href
  {https://doi.org/10.1039/C5SM01976A} {\bibfield  {journal} {\bibinfo
  {journal} {Soft Matter}\ }\textbf {\bibinfo {volume} {12}},\ \bibinfo {pages}
  {508} (\bibinfo {year} {2016})}\BibitemShut {NoStop}%
\bibitem [{\citenamefont {Bera}\ \emph {et~al.}(2018)\citenamefont {Bera},
  \citenamefont {Carrier}, \citenamefont {Backus}, \citenamefont {Bonn},
  \citenamefont {Shahidzadeh},\ and\ \citenamefont {Bonn}}]{Bera2018}%
  \BibitemOpen
  \bibfield  {author} {\bibinfo {author} {\bibfnamefont {B.}~\bibnamefont
  {Bera}}, \bibinfo {author} {\bibfnamefont {O.}~\bibnamefont {Carrier}},
  \bibinfo {author} {\bibfnamefont {E.~H.~G.}\ \bibnamefont {Backus}}, \bibinfo
  {author} {\bibfnamefont {M.}~\bibnamefont {Bonn}}, \bibinfo {author}
  {\bibfnamefont {N.}~\bibnamefont {Shahidzadeh}},\ and\ \bibinfo {author}
  {\bibfnamefont {D.}~\bibnamefont {Bonn}},\ }\bibfield  {title} {\bibinfo
  {title} {Counteracting interfacial energetics for wetting of hydrophobic
  surfaces in the presence of surfactants},\ }\href
  {https://doi.org/10.1021/acs.langmuir.8b02874} {\bibfield  {journal}
  {\bibinfo  {journal} {Langmuir}\ }\textbf {\bibinfo {volume} {34}},\ \bibinfo
  {pages} {12344} (\bibinfo {year} {2018})}\BibitemShut {NoStop}%
\bibitem [{\citenamefont {Tadmor}\ \emph {et~al.}(2019)\citenamefont {Tadmor},
  \citenamefont {Baksi}, \citenamefont {Gulec}, \citenamefont {Jadhav},
  \citenamefont {N'guessan}, \citenamefont {Sen}, \citenamefont {Somasi},
  \citenamefont {Tadmor}, \citenamefont {Wasnik},\ and\ \citenamefont
  {Yadav}}]{Tadmor2019}%
  \BibitemOpen
  \bibfield  {author} {\bibinfo {author} {\bibfnamefont {R.}~\bibnamefont
  {Tadmor}}, \bibinfo {author} {\bibfnamefont {A.}~\bibnamefont {Baksi}},
  \bibinfo {author} {\bibfnamefont {S.}~\bibnamefont {Gulec}}, \bibinfo
  {author} {\bibfnamefont {S.}~\bibnamefont {Jadhav}}, \bibinfo {author}
  {\bibfnamefont {H.~E.}\ \bibnamefont {N'guessan}}, \bibinfo {author}
  {\bibfnamefont {K.}~\bibnamefont {Sen}}, \bibinfo {author} {\bibfnamefont
  {V.}~\bibnamefont {Somasi}}, \bibinfo {author} {\bibfnamefont
  {M.}~\bibnamefont {Tadmor}}, \bibinfo {author} {\bibfnamefont
  {P.}~\bibnamefont {Wasnik}},\ and\ \bibinfo {author} {\bibfnamefont
  {S.}~\bibnamefont {Yadav}},\ }\bibfield  {title} {\bibinfo {title} {Drops
  that change their mind: spontaneous reversal from spreading to retraction},\
  }\href {https://doi.org/10.1021/acs.langmuir.9b02592} {\bibfield  {journal}
  {\bibinfo  {journal} {Langmuir}\ }\textbf {\bibinfo {volume} {35}},\ \bibinfo
  {pages} {15734} (\bibinfo {year} {2019})}\BibitemShut {NoStop}%
\bibitem [{\citenamefont {Kirkwood}\ and\ \citenamefont
  {Riseman}(1948)}]{Kirkwood1948}%
  \BibitemOpen
  \bibfield  {author} {\bibinfo {author} {\bibfnamefont {J.~G.}\ \bibnamefont
  {Kirkwood}}\ and\ \bibinfo {author} {\bibfnamefont {J.}~\bibnamefont
  {Riseman}},\ }\bibfield  {title} {\bibinfo {title} {The intrinsic viscosities
  and diffusion constants of flexible macromolecules in solution},\ }\href
  {https://doi.org/10.1063/1.1746947} {\bibfield  {journal} {\bibinfo
  {journal} {J.~Chem.~Phys.}\ }\textbf {\bibinfo {volume} {16}},\ \bibinfo
  {pages} {565} (\bibinfo {year} {1948})}\BibitemShut {NoStop}%
\bibitem [{\citenamefont {Tanford}(1972)}]{Tanford1972}%
  \BibitemOpen
  \bibfield  {author} {\bibinfo {author} {\bibfnamefont {C.}~\bibnamefont
  {Tanford}},\ }\bibfield  {title} {\bibinfo {title} {Micelle shape and size},\
  }\href {https://doi.org/10.1021/j100665a018} {\bibfield  {journal} {\bibinfo
  {journal} {J.~Phys.~Chem.}\ }\textbf {\bibinfo {volume} {76}},\ \bibinfo
  {pages} {3020} (\bibinfo {year} {1972})}\BibitemShut {NoStop}%
\bibitem [{\citenamefont {Decker}\ \emph {et~al.}(1999)\citenamefont {Decker},
  \citenamefont {Frank}, \citenamefont {Suo},\ and\ \citenamefont
  {Garoff}}]{Decker1999}%
  \BibitemOpen
  \bibfield  {author} {\bibinfo {author} {\bibfnamefont {E.~L.}\ \bibnamefont
  {Decker}}, \bibinfo {author} {\bibfnamefont {B.}~\bibnamefont {Frank}},
  \bibinfo {author} {\bibfnamefont {Y.}~\bibnamefont {Suo}},\ and\ \bibinfo
  {author} {\bibfnamefont {S.}~\bibnamefont {Garoff}},\ }\bibfield  {title}
  {\bibinfo {title} {Physics of contact angle measurements},\ }\href
  {https://doi.org/10.1016/S0927-7757(99)00069-2} {\bibfield  {journal}
  {\bibinfo  {journal} {Colloids Surf. A}\ }\textbf {\bibinfo {volume} {156}},\
  \bibinfo {pages} {177} (\bibinfo {year} {1999})}\BibitemShut {NoStop}%
\bibitem [{\citenamefont {Frank}\ and\ \citenamefont
  {Garoff}(1995)}]{Frank1995}%
  \BibitemOpen
  \bibfield  {author} {\bibinfo {author} {\bibfnamefont {B.}~\bibnamefont
  {Frank}}\ and\ \bibinfo {author} {\bibfnamefont {S.}~\bibnamefont {Garoff}},\
  }\bibfield  {title} {\bibinfo {title} {Temporal and spatial development of
  surfactant self-assemblies controlling spreading of surfactant solutions},\
  }\href {https://doi.org/10.1021/la00011a027} {\bibfield  {journal} {\bibinfo
  {journal} {Langmuir}\ }\textbf {\bibinfo {volume} {11}},\ \bibinfo {pages}
  {4333} (\bibinfo {year} {1995})}\BibitemShut {NoStop}%
\bibitem [{\citenamefont {Jarosiewicz}\ \emph {et~al.}(2004)\citenamefont
  {Jarosiewicz}, \citenamefont {Czechowski},\ and\ \citenamefont
  {Jad{\.{z}}yn}}]{Jarosiewicz2004}%
  \BibitemOpen
  \bibfield  {author} {\bibinfo {author} {\bibfnamefont {P.}~\bibnamefont
  {Jarosiewicz}}, \bibinfo {author} {\bibfnamefont {G.}~\bibnamefont
  {Czechowski}},\ and\ \bibinfo {author} {\bibfnamefont {J.}~\bibnamefont
  {Jad{\.{z}}yn}},\ }\bibfield  {title} {\bibinfo {title} {The viscous
  properties of diols v. 1,2-hexanediol in water and butanol solutions},\
  }\href {https://doi.org/10.1515/zna-2004-0905} {\bibfield  {journal}
  {\bibinfo  {journal} {Z.~Naturforsch.~A}\ }\textbf {\bibinfo {volume}
  {59a}},\ \bibinfo {pages} {559} (\bibinfo {year} {2004})}\BibitemShut
  {NoStop}%
\bibitem [{\citenamefont {Karpitschka}\ and\ \citenamefont
  {Riegler}(2010)}]{Karpitschka2010}%
  \BibitemOpen
  \bibfield  {author} {\bibinfo {author} {\bibfnamefont {S.}~\bibnamefont
  {Karpitschka}}\ and\ \bibinfo {author} {\bibfnamefont {H.}~\bibnamefont
  {Riegler}},\ }\bibfield  {title} {\bibinfo {title} {Quantitative experimental
  study on the transition between fast and delayed coalescence of sessile
  droplets with different but completely miscible liquids},\ }\href
  {https://doi.org/10.1021/la1007457} {\bibfield  {journal} {\bibinfo
  {journal} {Langmuir}\ }\textbf {\bibinfo {volume} {26}},\ \bibinfo {pages}
  {11823} (\bibinfo {year} {2010})}\BibitemShut {NoStop}%
\bibitem [{\citenamefont {Smit}\ \emph {et~al.}(1990)\citenamefont {Smit},
  \citenamefont {Schlijper}, \citenamefont {Rupert},\ and\ \citenamefont {van
  Os}}]{Smit1990}%
  \BibitemOpen
  \bibfield  {author} {\bibinfo {author} {\bibfnamefont {B.}~\bibnamefont
  {Smit}}, \bibinfo {author} {\bibfnamefont {A.~G.}\ \bibnamefont {Schlijper}},
  \bibinfo {author} {\bibfnamefont {L.~A.~M.}\ \bibnamefont {Rupert}},\ and\
  \bibinfo {author} {\bibfnamefont {N.~M.}\ \bibnamefont {van Os}},\ }\bibfield
   {title} {\bibinfo {title} {Effects of chain length of surfactants on the
  interfacial tension: molecular dynamics simulations and experiments},\ }\href
  {https://doi.org/10.1021/j100381a003} {\bibfield  {journal} {\bibinfo
  {journal} {J.~Phys.~Chem.}\ }\textbf {\bibinfo {volume} {94}},\ \bibinfo
  {pages} {6933} (\bibinfo {year} {1990})}\BibitemShut {NoStop}%
\bibitem [{\citenamefont {Hu}\ \emph {et~al.}(1995)\citenamefont {Hu},
  \citenamefont {Xiao}, \citenamefont {Ogletree},\ and\ \citenamefont
  {Salmeron}}]{Hu1995}%
  \BibitemOpen
  \bibfield  {author} {\bibinfo {author} {\bibfnamefont {J.}~\bibnamefont
  {Hu}}, \bibinfo {author} {\bibfnamefont {X.-D.}\ \bibnamefont {Xiao}},
  \bibinfo {author} {\bibfnamefont {D.~F.}\ \bibnamefont {Ogletree}},\ and\
  \bibinfo {author} {\bibfnamefont {M.}~\bibnamefont {Salmeron}},\ }\bibfield
  {title} {\bibinfo {title} {Imaging the condensation and evaporation of
  molecularly thin films of water with nanometer resolution},\ }\href
  {https://doi.org/10.1126/science.268.5208.267} {\bibfield  {journal}
  {\bibinfo  {journal} {Science}\ }\textbf {\bibinfo {volume} {268}},\ \bibinfo
  {pages} {267} (\bibinfo {year} {1995})}\BibitemShut {NoStop}%
\bibitem [{\citenamefont {Wijshoff}(2018)}]{Wijshoff2018}%
  \BibitemOpen
  \bibfield  {author} {\bibinfo {author} {\bibfnamefont {H.}~\bibnamefont
  {Wijshoff}},\ }\bibfield  {title} {\bibinfo {title} {Drop dynamics in the
  inkjet printing process},\ }\href
  {https://doi.org/https://doi.org/10.1016/j.cocis.2017.11.004} {\bibfield
  {journal} {\bibinfo  {journal} {Curr.~Opin.~Colloid~Interface~Sci.}\ }\textbf
  {\bibinfo {volume} {36}},\ \bibinfo {pages} {20} (\bibinfo {year}
  {2018})}\BibitemShut {NoStop}%
\end{thebibliography}
\end{document}